\documentstyle[sevensterapj,sevensterpsfig]{sevensterart}
\def\shortauthors{\lefthead}
\def\shorttitle{\righthead}

\newcount\levelone    \levelone=0
\newcount\leveltwo    \leveltwo=0
\newcount\levelthree  \levelthree=0
\newcount\levelfour   \levelfour=0
\def\chaphead{}                             
\def\secno{\chaphead\the\levelone}
\def\subno{\chaphead\the\levelone.\the\leveltwo}
\def\subsubno{\chaphead\the\levelone.\the\leveltwo.\the\levelthree}
\def\subsubsubno{\chaphead\the\levelone.\the\leveltwo.\the\levelthree
                           .\the\levelfour}
\def\newsec{\advance\levelone by1 \leveltwo=0 \levelthree=0 \levelfour=0}
\def\newsub{\advance\leveltwo by1 \levelthree=0 \levelfour=0}
\def\newsubsub{\advance\levelthree by1 \levelfour=0}
\def\newsubsubsub{\advance\levelfour by1}

\def\absnarrower{\advance\leftskip by \abstractindent}

\newdimen\secskipamount  \secskipamount=1pt
\newdimen\subskipamount  \subskipamount=1pt
\newdimen\bottomtol \bottomtol=0.03\vsize
\def\secskip{\par \ifdim\lastskip<\secskipamount \removelastskip \fi
    \vskip 0pt plus \bottomtol \penalty-250
    \vskip 0pt plus -\bottomtol \relax
    \vskip\secskipamount plus3pt minus3pt}
\def\subskip{\par \ifdim\lastskip<\subskipamount \removelastskip \fi
    \vskip 0pt plus 0.5\bottomtol \penalty-150
    \vskip 0pt plus -0.5\bottomtol \relax
    \vskip\subskipamount plus2pt minus2pt}
\def\subsubskip{\par \ifdim\lastskip<\subskipamount \removelastskip \fi
    \vskip 0pt plus 0.5\bottomtol \penalty-150
    \vskip 0pt plus -0.5\bottomtol \relax
    \vskip\subskipamount plus2pt minus2pt \hskip 10pt}

\outer\def\unnumberedsectionbegin #1 #2\par {\secskip \noindent {{\bf  #1}
\dotfill #2}
    \nobreak \vskip 1pt \noindent}

\outer\def\sectionbegin #1 #2\par {\secskip \newsec \noindent {{\bf \secno\  #1}
\dotfill #2}
    \nobreak \vskip 1pt \noindent}
\outer\def\subsectionbegin #1 #2\par {\subskip \newsub {\subno\ {\rm #1} \hfill
#2}
    \nobreak \vskip 1pt \noindent}
\outer\def\subsubsectionbegin #1 #2\par {\subsubskip \newsubsub
    {\subsubno\ {\it #1} \hfill #2}
  \nobreak \vskip 1pt \noindent}
\newcount\eqnumber \eqnumber=0
\def\new{{\rm\chaphead\the\eqnumber}\global\advance\eqnumber by 1}

\newcount\fignumber \fignumber=0
\def\nfig{\chaphead\the\fignumber\global\advance\fignumber by 1}
\def\ntab{\chaphead\the\tabnumber\global\advance\tabnumber by 1}

\newcount\fononum \fononum=0
\def\nfn{\global\advance\fononum by 1}
\def\fonono{\the\fononum}

\def\bck{\hskip-0.35em}
\def\wisk#1{\ifmmode{#1}\else{$#1$}\fi}
\def\etal{{et al.$\,$}}

\def\twcc{colour--colour diagram}

\def\mum{\wisk{\mu}m}

\let\msun=\msol

\def\decdeg#1.#2 {\wisk{#1^{\,\rm o}\bck.\,#2}\ }
\def\decmin#1.#2 {\wisk{#1^{\,\prime}\bck.\,#2}\ }

\def\decsec#1.#2 {\wisk{#1^{\prime\prime}\hskip-0.42em.\hskip0.10em#2}\ }
\def\arcsec {\wisk{^{\prime\prime}}\ }
\def\kms{\wisk{\,\rm km\,s^{-1}\,}}                    

\def\oversim#1#2{\lower1.5pt\vbox{\baselineskip0pt \lineskip-0.5pt
     \ialign{$\mathsurround0pt #1\hfil##\hfil$\crcr#2\crcr\sim\crcr}}}
\def\gsim{\wisk{\mathrel{\mathpalette\oversim{>}}}} 
\def\lsim{\wisk{\mathrel{\mathpalette\oversim{<}}}} 

\def\eqref#1{\advance\eqnumber by -#1 \chaphead\the\eqnumber
           \advance\eqnumber by #1 }
\def\?{\eqref{1}}
\def\last{\advance\eqnumber by -1 {\rm\chaphead\the\eqnumber}\advance
     \eqnumber by 1}
\def\eqnam#1{\xdef#1{\chaphead\the\eqnumber}}

\def\nfig{\chaphead\the\fignumber\global\advance\fignumber by 1}
\def\anfig{\global\advance\fignumber by 1}

\def\ntab{\chaphead\the\tabnumber\global\advance\tabnumber by 1}
\def\antab{\global\advance\tabnumber by 1}
\def\nfiga#1{\chaphead\the\fignumber{#1}\global\advance\fignumber by 1}
\def\rfig#1{\advance\fignumber by -#1 \chaphead\the\fignumber
            \advance\fignumber by #1}
\def\fignam#1{\xdef#1{\chaphead\the\fignumber}}
\def\tabnam#1{\xdef#1{\chaphead\the\tabnumber}}
\def\aa#1 #2 {, {A\&A,}{ #1, #2} }
\def\aal#1 #2 {, {A\&A,}{ #1, L#2}\ }
\def\aas#1 #2 {, {A\&AS,}{ #1, #2} }
\def\aj#1 #2 {, {AJ, }{#1, #2}\ }
\def\apj#1 #2 {, {ApJ, }{#1, #2}\ }
\def\apjl#1 #2 {, {ApJ, }{#1, L#2 }\ }
\def\apjs#1 #2 {, {ApJS, }{#1, #2}\ }
\def\araa#1 #2 {, {ARA\&A, }{#1, #2}\ }
\def\mnras#1 #2 {, {MNRAS, }{#1, #2}\ }
\def\bargal#1 { {1996, In: {Buta, R., Crocker, D., Elmegreen, B.~(eds.)
      Barred Galaxies, PASPC 91, San Francisco,} p. #1 }}
\def\thrss#1 { {1999, In: Gibson, B., Axelrod, T., Putman, M.~(eds.)
     The Third Stromlo Symposium: The Galactic Halo,
     PASPC 165, San Francisco, p. #1 }}

\newcount\fignumber \fignumber=1
\newcount\eqnumber \eqnumber=1
\newcount\tabnumber \tabnumber=1

\def\Sct{Section\ }
\def\Eqt{Equation\ }
\def\Fg{Fig.~}

\def\TDSI{1}
\def\ESO{3}

\def\SPI{A1}
\def\SPM{A2}
\def\SPH{A3}
\def\SBM{A4}
\def\SCI{A5}
\def\SPS{A6}
\def\SIS{A7}
\def\SES{A8}

\def\HSI{1}
\def\HSM{2}
\def\MSXA{3}
\def\IRSA{4}
\def\VEXA{5}
\def\IRSB{6}
\def\NIR{7}
\def\HFA{8}
\def\HKKT{9}
\def\SFRS{10}
\def\VDVA{11}
\def\VDVB{12}
\def\BOL{13}
\def\SOSI{14}

\def\bibitm{\bibitem{}}

\begin{document}

\title{OH--selected AGB and post--AGB objects.\\
   I.  Infrared and maser properties}

\author{Maartje N.~Sevenster\altaffilmark{1}}

\altaffiltext{1}{sevenste@strw.leidenuniv.nl}

\affil{MSSSO/RSAA, Cotter Road, Weston ACT 2611, Australia}

\shortauthors{M.~Sevenster}
\shorttitle{Infrared and maser properties of an OH sample}

\begin{abstract}
Using 766 compact objects found in a systematic survey of the
galactic Plane in the 1612--MHz masing OH line, new light is cast
on the infrared properties of evolved stars on the
thermally--pulsing asymptotic giant branch and beyond.
The usual mid--infrared selection criteria for post--AGB, based on IRAS 
colours, largely fail to distinguish early post--AGB stages.
A two--colour diagram from much narrower--band MSX flux densities,
with bimodal distributions, provides a better tool to do the latter. 
Four mutually consistent selection 
criteria for OH--masing red proto--planetary nebulae are given, as well
as two for early post--AGB masers and one for all post--AGB
masers including the earliest ones.
All these criteria miss a group of blue, high--outflow 
post--AGB sources with 60--\mum\ excess; these will be discussed 
in detail in Paper II.
The majority of post--AGB sources show
regular double--peaked spectra in the OH 1612--MHz line, with fairly 
low outflow velocities, although
the fractions of single peaks and irregular spectra may vary with age
and mass. The OH flux density shows a fairly regular relation with
the stellar flux and the envelope optical depth, with the maser efficiency
increasing with IRAS colour $R_{21}$. The OH flux density is 
linearly correlated with the 60--\mum\ flux density.
\end{abstract}


\section{Introduction}

Maser emission in the satellite line of groundstate 
OH at 1612.231 MHz occurs in a variety of stellar objects.
Star--forming regions (SFR), super giants, visible (Miras)
and invisible (OH/IR) oxygen--rich stars at the tip of
the asymptotic giant branch (AGB) and post--AGB transition 
objects or even young planetary nebulae can all show 
maser emission at 1612 MHz. Spectral
shapes and variability are thought to differ for the different types of 
object (see eg. te Lintel Hekkert \& Chapman 1996).

Objects are often selected as AGB or post--AGB 
candidates by the two colours from their
flux densities at 12 \mum , 25 \mum\ and 60 \mum\ measured by 
IRAS (InfraRed Astronomical Satellite).
(eg.~te Lintel Hekkert \etal1991; Lewis 1994 and references therein). 
In the resulting two--colour diagram, a ``main sequence'' for 
oxygen--rich AGB stars can
be identified, where the location of an object is determined
by its age, initial mass and possibly
metallicity (see eg.~Bedijn 1987; Likkel 1989).
This sequence is called the ``evolutionary'' 
sequence (van der Veen \& Habing 1988).
To test the usual selections, especially of post--AGB objects,
and potentially find less ambiguous criteria, in this paper the approach
is, for the first time, 
from an OH--selected sample, finding its IR properties,
rather than finding the OH properties of an IR--selected 
sample. Moreover, IRAS colours are combined with colours at
shorter infrared wavelengths from the Midcourse Space Experiment (MSX)
and 2--Micron All Sky Survey (2MASS).

The sample consists of 766 compact, OH--masing
sources{\footnote {A self--explanatory archive of all OH and IR data 
for all objects can be downloaded from http://www.mso.anu.edu/$\sim$msevenst}}
in the 
galactic Plane (for all details see Sevenster \etal 1997a,1997b,2001).
The IRAS (12,25,60,100 \mum ), MSX (4,8,12,15,21 \mum )
and 2MASS identifications (J,H,K) were taken
from the respective public databases. Matches were
found for 587 (IRAS, within 3$\sigma$ error ellipse), 
494 (MSX, within 5\arcsec) 
and 194 (2MASS, within 3\arcsec ) objects, respectively. 
Note that
the 2MASS mission has not yet covered the entire section of the plane
covered by the OH sample.
For many sources that do have a valid positional identification,
several or most of the flux densities may not have been measured
properly. For instance, out of the 587 IRAS identifications,
only 240 have well--determined 12--\mum\ , 25--\mum\ \'and 60--\mum\ flux
densities (quality flag 2 or 3).
The relatively small percentage of IRAS identifications
is caused largely by the confusion at the very low latitudes
of our survey ($|b| < 3^{\circ}.25$).
For the same reason, the flux--density
measurements, especially at 60 \mum\ and 100 \mum,  are likely
to be somewhat overestimated even if they are not
flagged as ``upper limits'' in the IRAS data base.
With the exception of a few likely supergiants, none of 
the sources could be identified with an optical counterpart.
The combination of low latitude and OH--based selection means 
that lowest--mass AGB stars ($<$ 1\msun ) may be underrepresented
in the sample.

\begin{figure*}
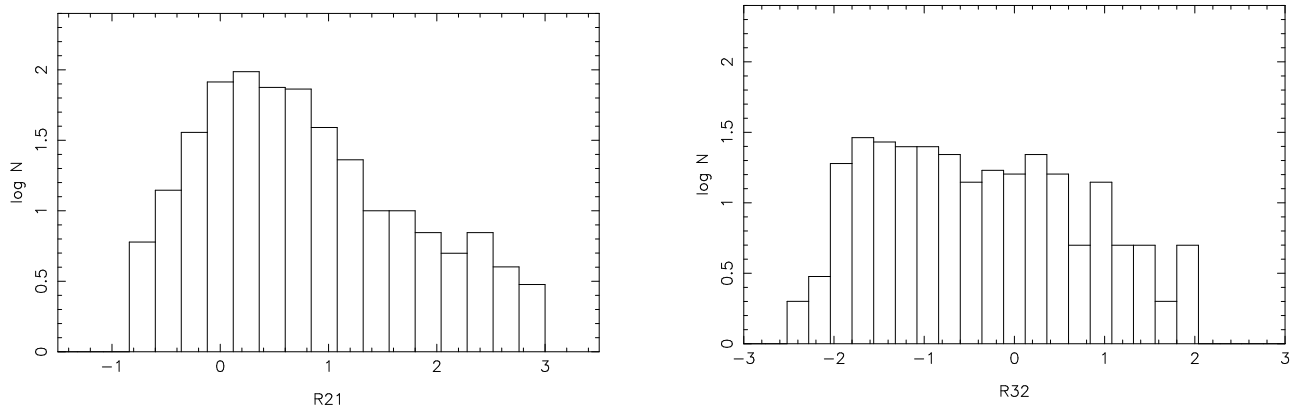

\fignam\HSI
\anfig
\psfig{figure=Sevenster1.fig1a.ps,angle=270,width=8truecm}
\vskip -5.5truecm
\hskip 9truecm{
\psfig{figure=Sevenster1.fig1b.ps,angle=270,width=8truecm}}
\figcaption{
The histograms of the two main IRAS colours ($R_{21}=[12-25]$ 
and $R_{32}=[25-60]$) for all OH--selected sources with 
well--determined flux--density measurements for all three
bands involved.
}
\end{figure*}

The aim of this analysis is to find generic differences 
between post--AGB and AGB objects, both in infrared and maser properties,
and signs of post--AGB evolution.
We will define a post--AGB stage as 
the phase after the last thermal pulse, when variability has ceased. 
The term proto--planetary nebula (PPN) will indicate a slightly
later stage, when the mass--loss rate has dropped by orders
of magnitude and the near--infrared or even optical
emission is becoming much stronger with respect to the mid--infrared.
Sometimes such objects are called OHPN, as they may
already show radio continuum (PN), but still harbour masers (OH).
However, as long as an object shows OH--maser emission, especially
at 1612 MHz as our sample does, it is per definition not
a full--blown PN (Zijlstra \etal1989; Lewis 1989).

\begin{figure*}
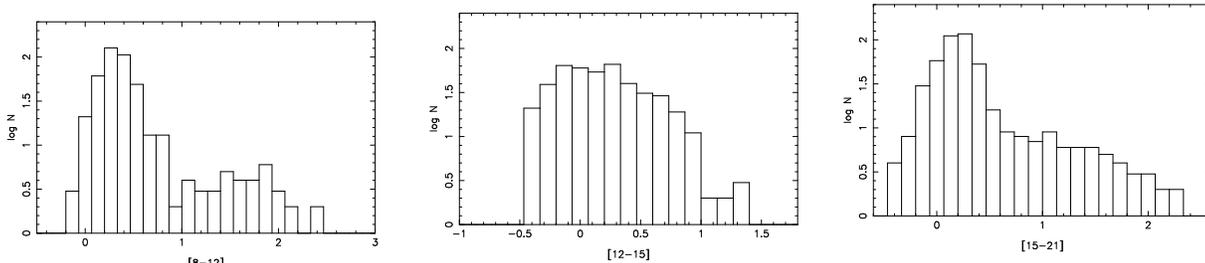

\fignam\HSM
\anfig
\psfig{figure=Sevenster1.fig2a.ps,angle=270,width=5truecm}
\vskip -3.5truecm
\hskip 5.5truecm{
\psfig{figure=Sevenster1.fig2b.ps,angle=270,width=5truecm}}
\vskip -3.5truecm
\hskip 11truecm{
\psfig{figure=Sevenster1.fig2c.ps,angle=270,width=5truecm}}
\figcaption{
The histograms for the three main MSX colours, for all 
sources with well--determined flux densities in both bands
of each colour. The $[8-12]$ and $[15-21]$ distributions
look very different from the $[12-15]$ distribution, as well
as from the distributions in \Fg\HSI , and
seem bimodal with minima around 0.9 and 0.7, respectively.
}
\end{figure*}

In \Sct 2, we will 
introduce an MSX two--colour diagram.
In \Sct 3, we discuss the role of the OH outflow velocity and
the spectral shapes of different types of objects and in \Sct 4 the
near--infrared properties are added to the discussion. In \Sct 5 some
selection criteria for late-- versus early post--AGB objects are
summarized and evidence from known sources presented, then 
in \Sct 6 a new expression for OH 1612--MHz maser efficiency in
(post--) AGB objects is given. 
Conclusions are presented in \Sct 7.

\begin{figure*}
\fignam\MSXA
\anfig
\psfig{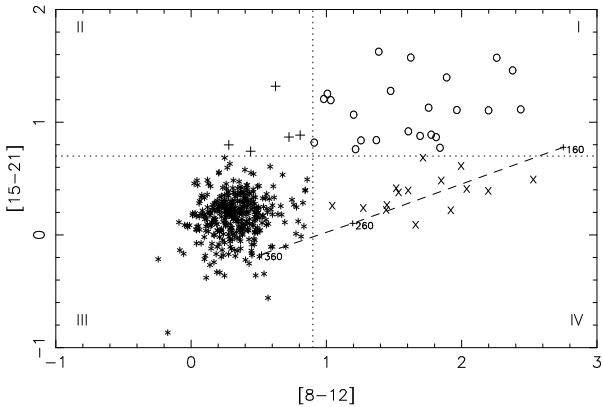}
\figcaption{
A two--colour diagram for MSX colours, split into four
quadrants according to the apparent minima in the bimodal
distributions shown in \Fg\HSM . Objects are plotted with 
different symbols accordingly (circles in the first
quadrant, plusses in the second, etc.), that will be used to relate
their locations in other diagrams to their MSX colours.
The dashed line shows the colours for black--body
radiation at various temperatures (tags).
}
\end{figure*}

\section{Distributions of mid--infrared colours}

All mid--infrared (MIR) colours are defined as
$[{\rm a}-{\rm b}]=2.5\log(S_{\rm b}/S_{\rm a})$ with $S$ 
flux density in Jy and a,b wavelength in \mum .
For IRAS, the usual names $R_{21}\equiv [12-25]$ and 
$R_{32}\equiv [25-60]$ are used, as well as $R_{43} \equiv [60-100]$.
The 12--\mum\ bands of the two satellites are different : for 
MSX it covers 11.1 \mum\ to 13.2 \mum , for IRAS 8.5 \mum\ to
15.5 \mum . This means, most importantly, that the 9.7--\mum\ silicate
feature is in the MSX 8--\mum\ band (6.8 \mum\ to 10.8 \mum ) 
whereas it is in the IRAS 12--\mum\ band. Only in very cold 
material may the silicate feature shift into the MSX 12--\mum\ band
(Zhang \& Kwok 1990). The MSX 15--\mum\ band is almost
entirely included in the IRAS 12--\mum\ band and
similarly the MSX 21--\mum\ band is
almost entirely included in the IRAS 25--\mum\ band.

\begin{figure*}
\fignam\IRSA
\anfig
\psfig{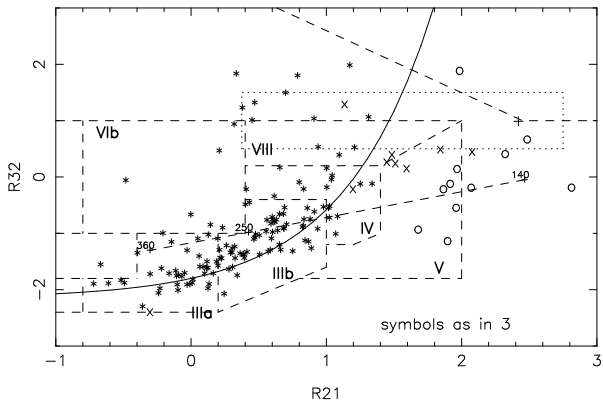}
\figcaption{
The two--colour diagram for IRAS colours, with symbols as
defined by the quadrants in \Fg\MSXA . The solid curve indicates
the evolutionary sequence and the dashed boxes the regions as
defined by van der Veen \& Habing (1988). The dashed line 
separates roughly the red corner 
of OH--emitting SFRs (Braz \& Sivagnanam 1987; see \Sct 5.3 ).
(Non--OH--emitting SFRs would be found to the left of this region
at $R_{32}>$2.) The dotted rectangle is the region classified as 
``bipolar outflow'' region by Zijlstra \etal(2001).
Note that compared to \Fg\MSXA\ there are far fewer objects, as now
the IRAS colours, as well as the MSX colours,
have to be well--defined as well as the MSX colours.
}
\end{figure*}

In \Fg\HSI\&\HSM , histograms are shown for five MIR colours.
The distributions for
$[8-12]$ and $[15-21]$ immediately stand out as curiously bimodal
when compared to the other three, with
divisions at $\sim$0.9 and $\sim$0.7, respectively.
To follow this up, we use those two colours for the MSX 
two--colour diagram. All sources with these MSX colours well--defined are
plotted in \Fg\MSXA .
The bulk of the sources is in the bluest quadrant, with little
spread in colour, but a second group is at much 
redder $[8-12]$ with a large spread in $[15-21]$.

\begin{figure*}
\fignam\VEXA
\anfig
\psfig{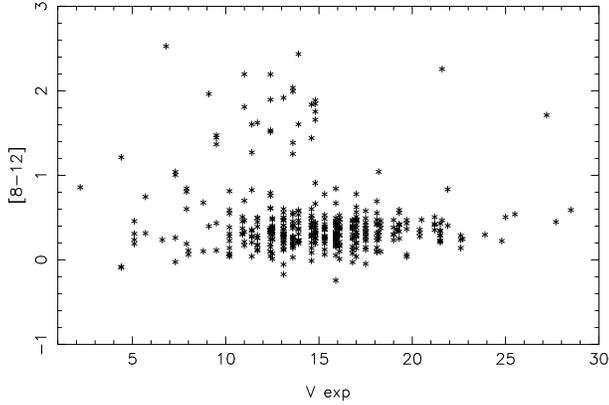}
\figcaption{
The MSX $[8-12]$ colour plotted against OH outflow velocity. The 
general group of AGB sources
with $[8-12]$$\,\sim\,$0.4 ranges over all outflow
velocities.
There is a discrete group of sources with much
redder colour, indicating they are post--AGB and PPN sources.
Their outflow velocities are limited to the range 9--15\kms .
The same trend is seen -- less clearly -- for $R_{21}$ and $[15-21]$,
but not for any of the other IR colours.
}
\end{figure*}

In \Fg\IRSA , the traditional IRAS \twcc\ is shown, with the
evolutionary sequence and the regions as laid out by
van der Veen \& Habing (1988) and symbols defined by MSX colour.
The circles 
are found in the region where one typically expects late
post--AGB objects by IRAS colours (region V and further;
Bedijn 1987; van der Veen \& Habing 1988 (``non--variable OH/IR stars'')).
The bulk of the main MSX group is along the 
evolutionary sequence, except for a surprisingly large
group in the left part of region VIII and above. 
The crosses are found mostly straddling the border between V and VIII,
below the ``bipolar outflow'' region, defined recently 
by Zijlstra \etal(2001). 
A solitary `+' is right on the edge of where one expects to
find OH--emitting SFRs (Braz \& Sivagnanam 1987); in this
region there is strong overlap between PPN and SFRs.

The bimodality of the two MSX--colour distributions suggests
the existence of at least two intrinsically different types
of objects in the sample. Combining this with what one
already understands from the traditional IRAS \twcc\ (van 
der Veen \& Habing 1988), it seems that the transition from 
blue ($<$0.9) to red ($>$0.9) $[8-12]$ coincides with the transition
off the AGB.
The transition from blue ($<$0.7) to red ($>$0.7) $[15-21]$
could well indicate a next evolutionary transition, further
away from the evolutionary sequence (\Fg\IRSA ), such as the 
drop in mass--loss rate and start of the fast PPN 
wind (cf.~van Hoof \etal1997).
Hence, the MSX diagram possibly distinguishes between different
post--AGB stages and provides an easier tool to separate 
post--AGB stars from AGB stars, but does not distinguish between
different AGB objects like the IRAS diagram with its evolutionary
sequence. The two diagrams should be used in combination.

\begin{figure*}
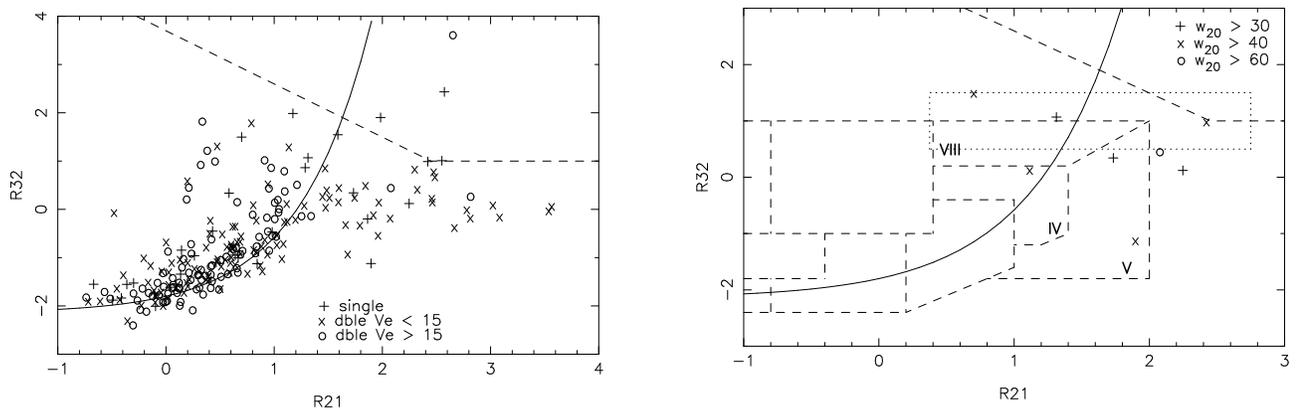

\fignam\IRSB
\anfig
\psfig{figure=Sevenster1.fig6a.ps,angle=270,width=8truecm}
\vskip -5.5truecm
\hskip 9truecm{
\psfig{figure=Sevenster1.fig6b.ps,angle=270,width=8truecm}}
\figcaption{
As \Fg\IRSA , with symbols according to outflow velocity. 
The extremely red single--peaked sources (`+') in the top panel
are most likely to be SFRs. The PPN region is
populated mostly by sources with outflow velocities below 15\kms\ (as
derives from \Fg\IRSA\&\VEXA ).
Interesting is the region to the left of the evolutionary
sequence, with $0<R_{32}<2$, where sources have mostly
outflow velocities {\it higher} than 15\kms . 
Before (\Fg\MSXA\& \IRSA), these
sources were found to have very blue MSX colours. 
Irregular sources (\Sct 3) are found in and around the bipolar--outflow
region (lower panel).
The spectra of the irregular sources are shown in \Fg\SIS .
}
\end{figure*}

\section{The role of the OH--shell outflow velocity}

The outflow velocity $V_{\rm exp}$ of a typical 
OH maser spectrum is thought to be related
to the luminosity (age, initial mass) and 
the metallicity (age, galactic coordinates) of the central star, via

\eqnam\ELV
$$L_{\ast} = V_{\rm exp}^4 Z^{-2}
\eqno(\new) $$

\noindent
according to van der Veen (1989; see also Habing \etal1994).
Not all OH maser spectra are regular and double--peaked,
so the outflow velocity is not always well-defined. 
Specifically, some suspected post--AGB objects may have
spectra ranging over almost 100 \kms\ (e.g. spectrum 2 in \Fg\SIS ;
see also te Lintel Hekkert \& Chapman 1996), 
but be listed as single--peaked with just the peak velocity, 
whereas similar objects do have an associated outflow velocity
(\Fg\SPI (4$^{\rm th}$--last)).
Such single--peaked objects are intrinsically different from
star-forming regions, that commonly -- especially with sufficient
spatial resolution -- have 
just one very narrow bright peak in their 1612--MHz spectrum, single--peaked
PN (\Fg\SPI (13)) or single--detected AGB stars, that simply
have one of the peaks missing due to detection limits (cf.~\Fg\SPI (15)).
Hence, selecting $V_{\rm exp}$=0\kms\ results in a hotchpotch of objects.

\begin{figure*}
\fignam\NIR
\anfig
\psfig{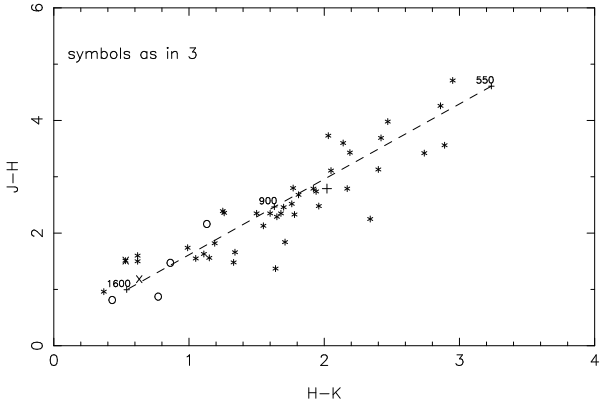}
\figcaption{
The 2MASS two--colour diagram (\Sct 4) with 
symbols defined by MSX colours (\Fg\MSXA ). 
The dashed line connects black--body colours for several temperatures.
For J$-$H$\,<\,$0.4, Garcia--Lario \etal(1997) find exclusively planetary
nebulae (PNe can be found at redder colours). None of our OH--masing
sources are that blue, 
which would agree with the idea that masers, especially the
1612--MHz OH maser, cannot exist
in the environment of a truely ionized, high--wind nebula.
}
\end{figure*}

We want to find out what OH outflow velocity and spectral ``shape''
tell us about the evolutionary stage of an object.
In \Fg\VEXA , $[8-12]$ is plotted against outflow velocity.
A population of red sources is clearly separated from the rest, with
outflow velocities between 9\kms\ and 15\kms . The separation is
at $[8-12]\sim$0.9, as before (\Fg\HSM\&\MSXA ); for $[15-21]$
a similar but less clear separation is found 
at $\sim$0.7 . Thus, interestingly, in the first and fourth
quadrant of \Fg\MSXA , all double--peaked sources are found
to be in this very limited range of outflow velocities.

\begin{figure*}
\fignam\HFA
\anfig
\psfig{figure=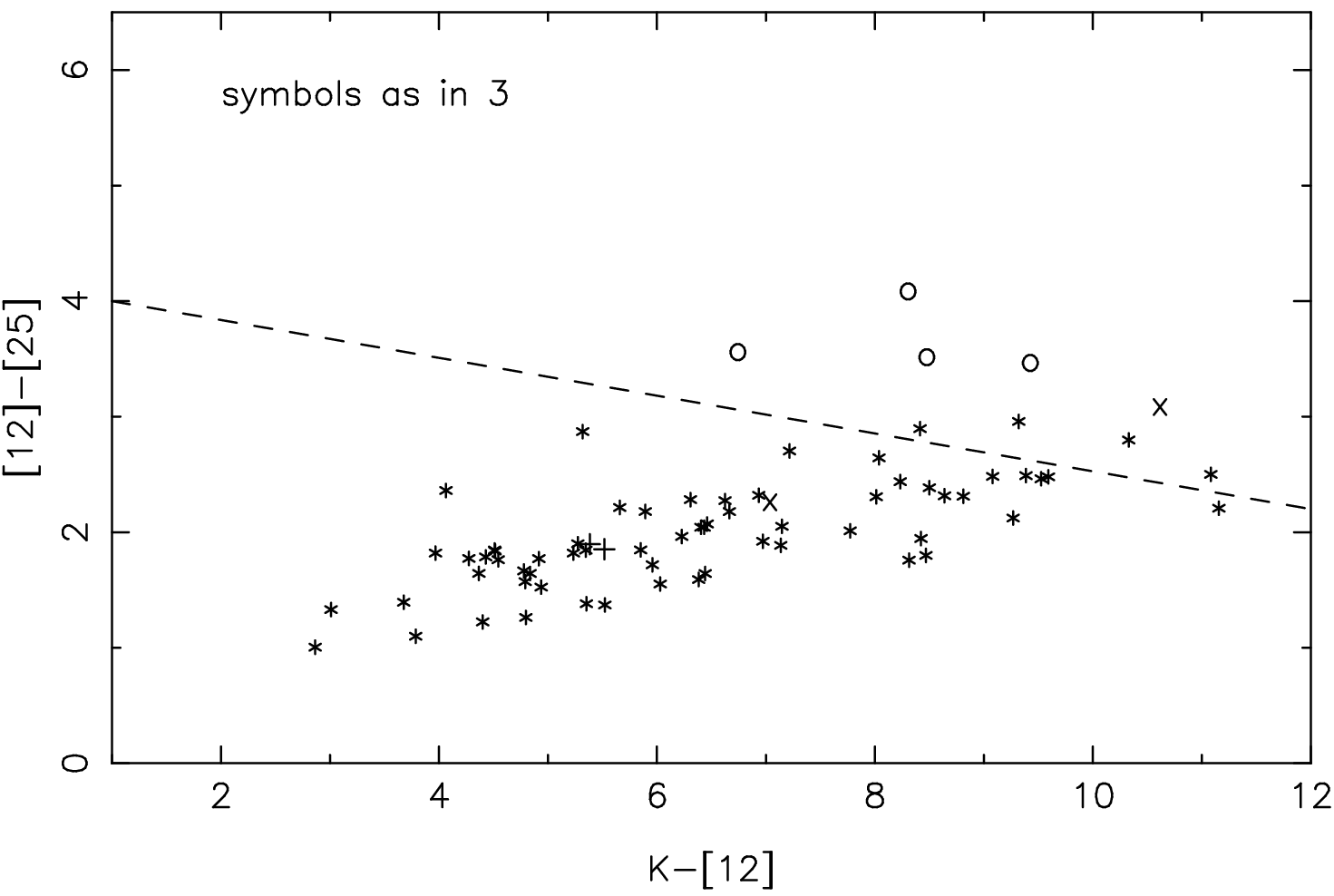,angle=270,width=8truecm}
\vskip -5.5truecm
\hskip 9truecm{
\psfig{figure=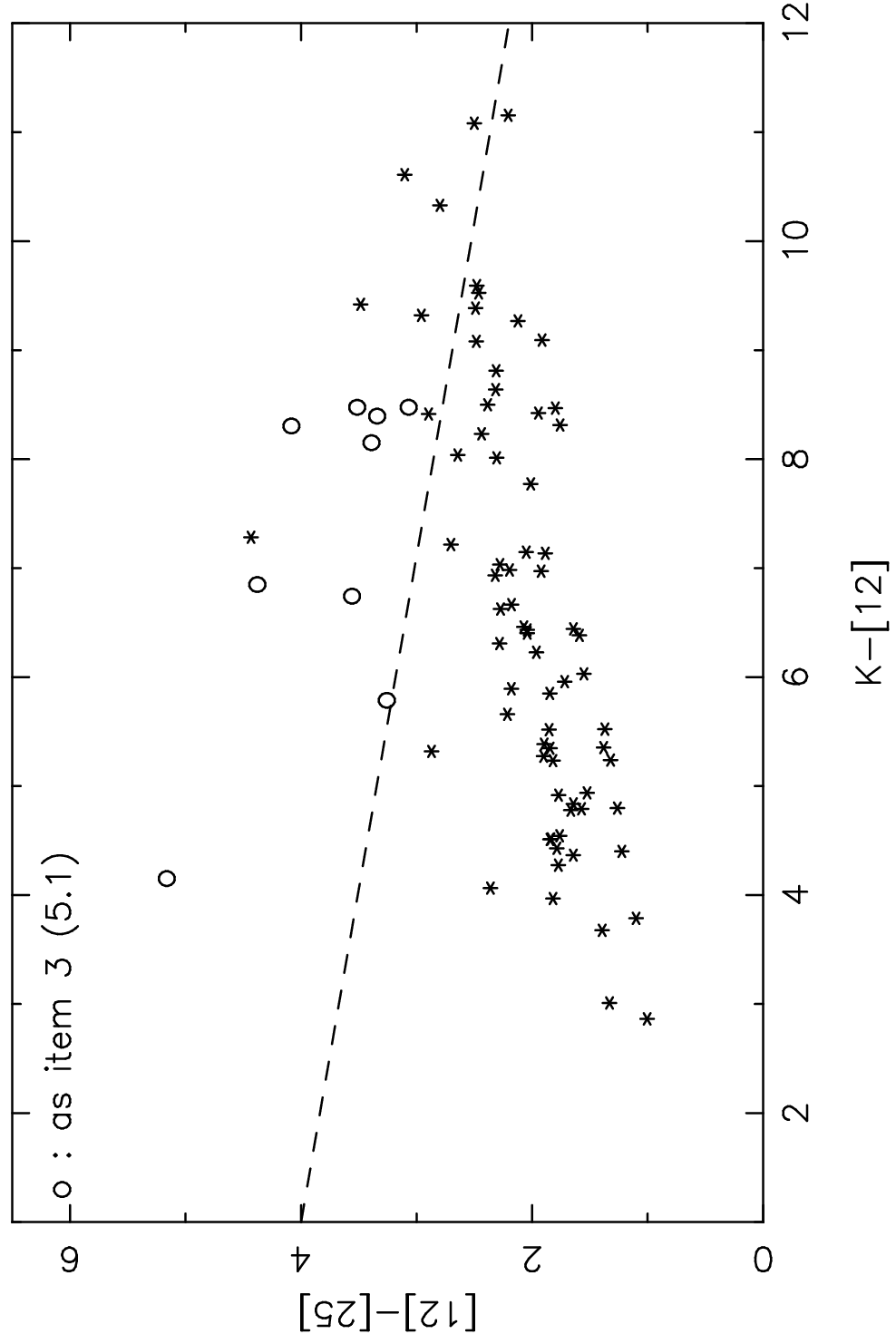,angle=270,width=8truecm}}
\figcaption{
The two--colour diagram using combined MIR and NIR colours as
defined in \Sct 4, following van Hoof \etal(1997).
{\bf a.} Symbols defined by MSX colour (\Fg\MSXA ).
{\bf b.} Symbols defined mainly by OH properties. The circles indicate
sources with outflow velocities between 9\kms\ and 15\kms ,
the sum of the two peak--flux densities larger than 1 Jy and 
$[15-21]$$>$0.7. They
also have $[15-21]$$\,>\,$0.7. Based on our previous findings, 
these sources are likely ``standard'' PPN sources.
Note that all circles have K$-[12]$$\,<\,$9, which is relevant in the
context of \Fg\SFRS .
}
\end{figure*}

In the IRAS \twcc\ (\Fg\IRSB a), region V and beyond
is populated mainly by lower--outflow velocity objects, as
follows from the previous two figures.
The single--peaked sources in the reddest corner of this
diagram are most likely SFRs.
On the evolutionary sequence, the spread in outflow
velocities is largest. We do not find the increase of outflow
velocity with $R_{21}$ for blue sources that is
reported in Chen \etal(2001), but our OH--selected sample
is probably too red to see this.
The higher outflow velocities prevail in the left--hand
sides of regions IV and VIII, where we noted relatively high 
numbers of sources before (\Sct 2). 
These sources may also be
post--AGB sources (van der Veen \& Habing 1988); they will be
discussed in Paper II (Sevenster 2002).

\begin{figure*}
\fignam\HKKT
\anfig
\psfig{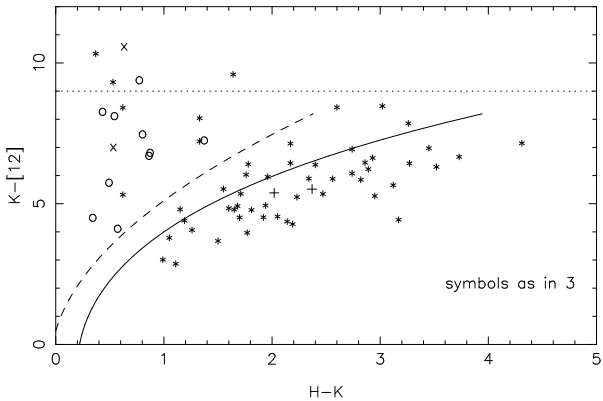}
\figcaption{
The two--colour diagram using combined MIR and NIR colours
as defined in \Sct 4, following van Loon \etal(1998).
The solid curve indicates their oxygen--rich AGB sequence; 
the dashed curve is our separation between AGB and post--AGB 
objects. The post--AGB sources, from quadrants I and IV (\Fg\MSXA ),
are all found above this line. Quite a few sources from quadrant III
are found in the post--AGB region, however, which suggest that
this
diagram might provide a tool to distinguish the earliest post--AGB stars.
Star--forming regions are found at K$-[12]$$\,>\,$9 (\Fg\SFRS ).
}
\end{figure*}

Post--AGB stars, typically selected from 
region V, are often connected to irregular, extreme--velocity
outflows (te Lintel Hekkert \& Chapman 1996).
However, looking at the OH spectra for these 
sources (\Fg\SPI ), they correspond to OH--selected 
sources with mostly normal, double--peaked, intermediate--outflow 
spectra (Table \TDSI ). 
The same is true for the higher--outflow sources to 
the left of the evolutionary sequence, although
the fraction of irregular spectra is higher (\Fg\SCI , Table \TDSI ).

This is not to say that there is no subset of 
irregular sources with reasonably confined properties.
In \Fg\SIS , the spectra are shown for all
sources from the full sample that have velocity
widths ($w_{20}$) much larger than the peak 
separation ($\Delta V \equiv 2V_{\rm exp}$). This criterion
is used as an indicator of non--spherical, irregular outflow.
Their IRAS colours (\Fg\IRSB b) clearly indicate 
a connection to the bipolar--outflow
region of Zijlstra \etal(2001). The actual velocity widths are up to
70 \kms . 

\begin{figure*}
\fignam\SFRS
\anfig
\psfig{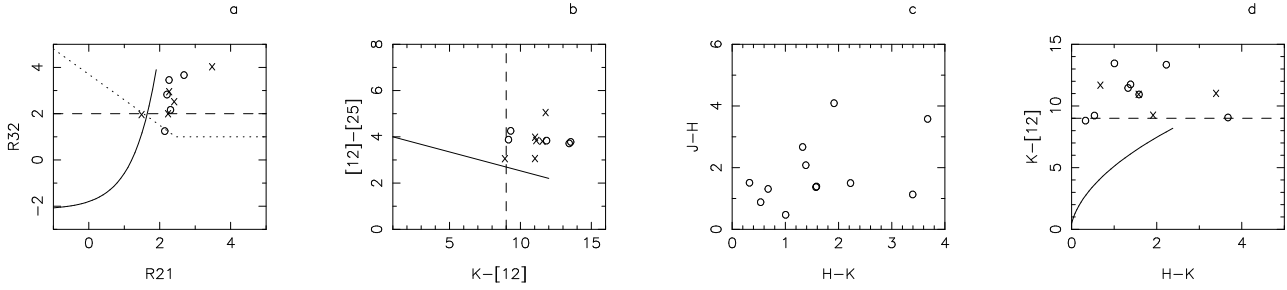}
\figcaption{
Two--colour diagrams
for a sample of known OH--masing SFRs (Testi \etal 1998).
Circles indicate reliable flux--density measurements and crosses
colours from flux--density upper limits.
In {\bf a}, the IRAS two--colour diagram 
is plotted, showing the ``OH--masing SFR''
region (\Sct 5.3).
In {\bf b}, the near--mid--infrared diagram shows that SFRs all have
K$-[12]$$\,>\,$9 but are above the post--AGB line; compare to \Fg\HFA .
Hence, the region 9$<$K$-[12]$$<$12 is shared by post--AGB stars
and SFRs, but SFRs do not venture into the ``PPN'' region (\Fg\HFA b). 
In {\bf c}, SFRs are seen to are spread over all NIR colours, hence
a good fraction has H$-$K higher than post--AGB objects. In {\bf d},
this means SFRs and post--AGB sources overlap
only in the region K$-[12]$$\,>\,$9 and H$-$K$\,<\,$1.5.
}
\end{figure*}

In summary, although very irregular OH spectra may always 
indicate a post--AGB host, the reverse may not be true.
Unless the strong reddening in $[8-12]$ and $R_{21}$ is
unrelated to the onset of post--AGB processes,
OH 1612--MHz spectra of post--AGB objects seem to be
mainly regular and show no indication
of irregular mass loss or increased outflow velocity, 
spherical nor bipolar. For the bulk of the red post--AGB objects, the
outflow velocities are even lower than average: between 9 \kms\ and
15 \kms .
Most likely, this indicates that the OH masers in general
disappear
before the changes of the central wind truly influence the envelope at
such large radii.

\begin{figure*}
\fignam\VDVA
\anfig
\psfig{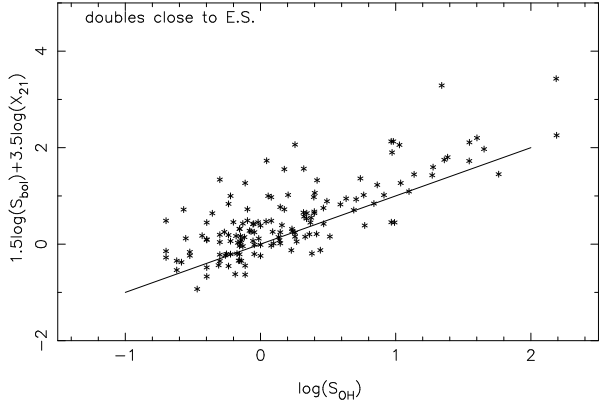}
\figcaption{
According to van der Veen (1989), the OH luminosity for
objects along the evolutionary sequence is proportional
to $L_{\ast}^2\, X_{21}^6$, with $X_{21}\equiv S_{25}/S_{12}$.
From this log--log plot, slightly different powers are found, 
1.5 for the stellar luminosity and 3.5 for the
flux--density ratio. 
}
\end{figure*}

\section{Near--infrared properties}

From the 2MASS database, J~(1.25 \mum ), H~(1.65 \mum ) and/or K~(2.17 \mum )
magnitudes were obtained
for 194 sources. Colours are used as uncorrected J--H and H--K.
The two--colour diagram is shown in \Fg\NIR .
Using the symbols from \Fg\MSXA , post--AGB
stars are where
one may expect them (e.g.~Garcia--Lario \etal 1997) but the
diagram can hardly be used as a diagnostic for any
particular type of object (see also \Fg\SFRS c).

\begin{figure*}
\fignam\VDVB
\anfig
\psfig{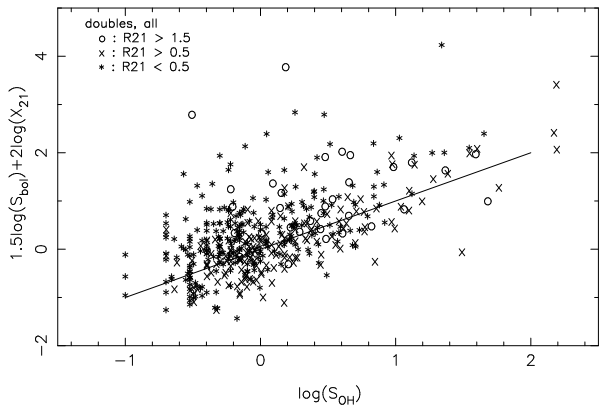}
\figcaption{
Same as \Fg\VDVA , including (double--peaked) sources away from 
the evolutionary sequence.
The power for $X_{21}$ is 2 in this case.
}
\end{figure*}

An alternative to the IRAS \twcc\ was proposed by
van Hoof \etal(1997), using $[12]\,-\,[25]$
and K$-[12]$ (\Fg\HFA ), with IRAS
magnitudes $[12]\equiv 2.5\log (59.5/S_{12})$ 
and $[25]\equiv 2.5\log (13.4/S_{25})$ 
following Oudmaijer \etal(1992).
The PPN symbols from \Fg\MSXA\ (circles) 
are exclusively at $[12]\,-\,[25]$\gsim 3.5.
In the bottom panel the selection is based on OH properties,
as well as $[15-21]$. The latter colour is used for the selection,
rather than $[8-12]$, because it may single out the 
further--evolved post--AGB sources (i.e.~PPNe).
Selecting $V_{\rm exp}$ between 9\kms\ and 15\kms\ (\Fg\VEXA )
and the sum of the OH flux densities of the two peaks larger than 1 Jy,
we find a sample (circles) that occupies the same region as 
the circles in \Fg\HFA (a). The dashed line in these diagrams 
is drawn to separate the circles from the other objects.

The post--AGB selection of \Fg\HFA (b), based largely on OH spectral
properties, gives {\it without further assumptions}
$[12]\,-\,[25]$$\,>\,$3 and K$-[12]$$\,<\,$9,
as well as $[8-12]$$\,>\,$0.9.
This once more confirms that the
sources in the first quadrant of \Fg\MSXA\ are 
``standard'' post--AGB sources, moving to higher $R_{21}$ in the IRAS 
two--colour diagram (Bedijn 1987;
van der Veen \& Habing 1988), and that these are mostly
regular and double--peaked (\Fg\SPH ).
A second mixed near--mid--infrared two--colour diagram is shown
in \Fg\HKKT . The red post--AGB group
separates clearly from the main oxygen--rich colour sequence
(see van Loon \etal(1998)).
However, some objects that are still in the third quadrant
of the MSX diagram -- and on the IRAS evolutionary
sequence -- are also above the dashed
curve and quite distinctly separated from the 
main colour sequence. This may indicate that they are in
the very first stages of post--AGB colour change.

\begin{figure*}
\fignam\BOL
\anfig
\psfig{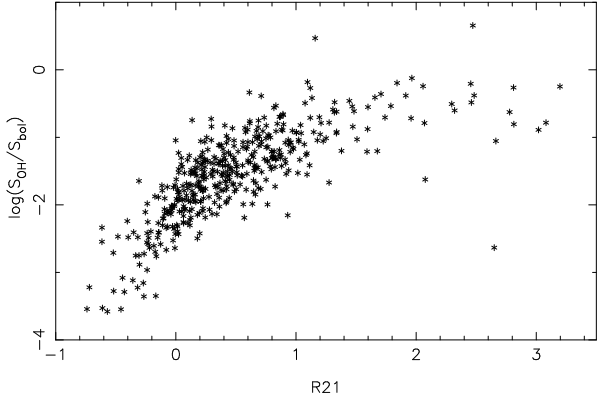}
\figcaption{
The maser ``efficiency'', $S_{\rm OH}/S_{\rm bol}$ 
(\Sct 6), is plotted logarithmically against IRAS colour $R_{21}$.
For the bluest sources, it is $\propto X_{21}^6$, for the reddest sources 
$\propto X_{21}^0$ (both $S_{\rm OH}$ and $S_{\rm bol}$ are almost
constant).
Again, only double--peaked sources are plotted to avoid SFRs.
}
\end{figure*}

In conclusion, the NIR magnitudes of objects are
useful to distinguish (post--) AGB evolutionary stages
primarily in combination with the 12--\mum\ and 25--\mum\ IRAS
flux densities.

\section{OH spectra of IR selected sources}

\subsection{Selecting post--AGB sources}

Summarizing the previous three sections, standard
``cold'' evolved OH PPN objects can thus be selected
virtually consistently in four ways.

\begin{itemize}

\item{(1) The traditional 
IRAS selection : $R_{32}$\lsim1.5 and $R_{21}$\gsim 1.4 (\Fg\IRSA ). 
Spectra are shown in \Fg\SPI .
}
\item{(2) From MSX colours : $[8-12]$$\,>\,$0.9 
and $[15-21]$$\,>\,$0.7 (\Fg\MSXA ).
Spectra are shown in \Fg\SPM .
}
\item{(3) From OH 1612--MHz properties :  $V_{\rm exp}=12\pm3$\kms\ and
$S_{\rm OH,blue}+S_{\rm OH,red}$$\,>\,$1 Jy (resolution $\sim$2\kms ), 
with $[15-21]$$\,>\,$0.7 (\Fg\HFA b).
Spectra are shown in \Fg\SPH .
}
\item{(4) With NIR : 
$[12]-[25] > 4.164-0.164\,({\rm K}-[12])$ and K$-[12]$$\,<\,$9
(\Fg\HFA a). This selection is
mostly a combination of the previous ones, but it can be used
for sources with known K magnitude and no MSX or 60--\mum\ association.
}

\end{itemize}

\noindent
As seen in all the corresponding figures, the selection
criteria, partly overlapping, are all entirely compatible, with 
just one or two sources breaking the rules.

\vskip .5truecm
\noindent
Early post--AGB sources are selected by :  

\begin{itemize}

\item{(5)  $ [15-21]$$\,<\,$0.7 and $[8-12]$$\,>\,$0.9 
(\Fg\MSXA ,\IRSA\&\HFA (a)), spectra in \Fg\SBM }

\item{(6)  $[12]-[25] > 4.164-0.164\,({\rm K}-[12])$ and
K$-[12]>$9 (\Fg\HFA a); these colours are shared by SFRs (\Fg\SFRS )}

\end{itemize}

\vskip .5truecm
\noindent
Post--AGB sources in general, including the very earliest, 
may be selected by :  

\begin{itemize}

\item{(7)  (H$-$K) + 0.05 $<$ 0.03 ((K$-$[12])+0.807)$^2$,
where early and late post--AGB objects can be distinguished 
by $[15-21]$ smaller or larger than 0.7 or even 
by $[8-12]$ smaller or larger than 0.9, respectively (\Fg\HKKT , \Fg\MSXA ).
}

\end{itemize}

\vskip .5truecm
\noindent
A group of blue possible post--AGB stars is found at :

\begin{itemize}

\item{(8)  $R_{32}$$\,>\,$$-$0.2 and $R_{21}$$\,>\,$0.2, 
to the left of the evolutionary sequence 
($R_{32} > -2.15 + 0.35\,\exp(1.5\,R_{21})$). 
}

\end{itemize}

\noindent
These are probably high--mass post--AGB 
sources (see Paper II)). 
Non--spherical--outflow objects (\Fg\SIS\&\IRSB b)
are likely to be found roughly in the region defined
by Zijlstra \etal(2001).

\tabnam\TDSI
\begin{deluxetable}{crrrrrrrrrrr}
\tablecaption{Fractions (in \%) of double (D), single (S)
and irregular (I) spectra for several post--AGB selections
plus comparison PN and AGB samples
(judged by eye from the figures in appendix A).
The first four rows correspond to selections
1,8,2 and 5 in \Sct 5.1, respectively}
\tablewidth{0pt}
\tablehead{
\colhead{ } &
\colhead{D } &
\colhead{S } &
\colhead{I } &
\colhead{comment }
}
\startdata
Right & 75 & 10 & 15 & iras (low--mass) \\
Left & 65 & 7  & 28 & iras (high--mass) \\
Q1 & 61 & 17 & 22 & msx (old) \\
Q4 & 75 & 0  & 25 & msx (young) \\
PN & 75 & 10  & 15 & simbad (\Sct 5.2) \\
AGB & 80 & 5  & 15 & iras (on evol.~seq.) \\
\enddata
\end{deluxetable}

\begin{figure*}
\fignam\SOSI
\anfig
\psfig{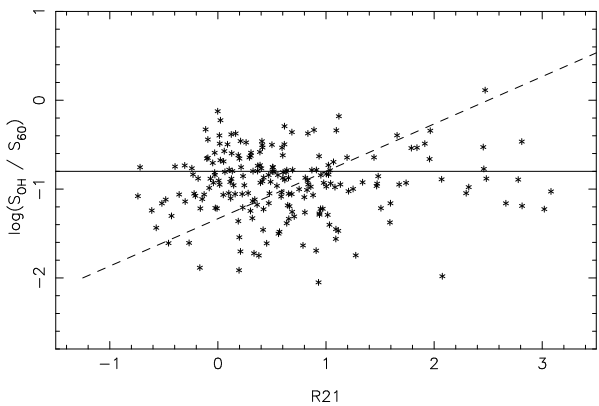}
\figcaption{
The ratio of OH--peak flux density and 60--\mum\ flux density
plotted logarithmically against IRAS colour $R_{21}$ for
all double--peaked sources.
Of all IRAS flux densities, $S_{60}$
correlates most strongly, almost linearly, with $S_{\rm OH}$.
The $S_{60}$ contains the 53--\mum\ pumping line.
The dashed line is the relation found by Chen \etal(2001)
for $S_{\rm OH}/S_{35}$, with $S_{35}$ the interpolated
flux density at the
35--\mum\ pumping line. 
}
\end{figure*}

\subsection{Known PNe in the sample}

Taking all the sources from our OH sample
that are associated with a ``Planetary Nebula'' within 
5\arcsec\ according to SIMBAD,
10 out of 16 do satisfy the OH criteria of outflow velocity 
and flux density. 
Four of the 16 have a 2MASS association, with K$-[12]$$\,<\,$9.4 and
above the lines in \Fg\HFA\&\HKKT .

Fourteen have an IRAS association, with 8 
in the traditional PPN region to the right
of the evolutionary sequence and 1 to the left,
with $V_{\rm exp}$$\,>\,$15 \kms\ and very blue
MSX colours. 
Of the 12, out of 16, sources with MSX association, 5 are in the first
quadrant (\Fg\MSXA ).

Spectra of all 16 sources are given in \Fg\SPS .
Even of this group of selected ``Planetary Nebulae'', 75\% have regular,
double--peaked spectra.
Even two known ``bipolar planetary nebulae'' are in this group
of 16 : OH353.844+02.984 
(\Fg\SPS (3); Kwok \etal1996) and OH348.813$-$02.840 
(\Fg\SPS (6); Hrivnak \etal1999).
The latter satisfies all our selection criteria for which 
its properties are known, including slow, regular outflow.

In summary, sources classified as ``planetary nebulae'' in 
fact cover a range of post--AGB evolutionary stages.
As we noted in the caption of \Fg\NIR , 
a source that still harbours 1612--MHz emission is unlikely to be
a full--blown PN. It would be good practice to name such objects
PPN or OHPN (see Introduction).

\subsection{Colours of SFR sources}

It is hard to define clear--cut selection criteria to find
SFRs in an OH sample, especially to tell them apart from PPN.
A sample of known OH--masing SFRs with NIR data, from the literature 
(Testi \etal 1998), is used to check their infrared behaviour
with respect to post--AGB stars. In \Fg\SFRS , several two--colour diagrams
are given for comparison with the diagrams in the rest of the 
paper. 
The dotted line in \Fg\SFRS (a) outlines the OH--masing SFR region, slightly 
adapted from Braz \& Sivagnanam (1987), who claim that non--OH--masing
SFRs are found to the left of this region, at lower $R_{21}$ but
similar $R_{32}$.

Interestingly, in K$-[12]$ versus $[12]-[25]$ (\Fg\SFRS b), the SFRs
lie above the empirical line drawn in \Fg\HFA, but at higher
K$-[12]$ than the PPN sources. The selected sources in 
\Fg\HFA b and the SFRs in \Fg\SFRS b are perfectly 
separated, although there are some 
earlier post--AGB sources in the 
region K$-[12]$$\,>\,$9 (\Fg\HFA a).
In \Fg\SFRS d, the SFRs are also in the post--AGB region, 
again at higher K$-[12]$ and mostly higher H$-$K (\Fg\SFRS c).
Three SFRS with a reliable MSX identification with
(some) well--determined flux densities are all in quadrant II.

\section{Maser--infrared correlation}

In order to determine a maser ``efficiency'' in terms
of OH flux density as a function of stellar flux, we
derive the bolometric flux, 
$S_{\rm bol}\equiv S_{12}*BC_{12}$, using the 
bolometric correction 

\eqnam\EBC
 $$ BC_{12} = 0.7 + 2.9\exp(-3.0\,R_{21})+0.9\exp(0.7\,R_{21})
\eqno(\new) $$

\noindent
from van der Veen \& Breukers (1989). 
If all the stellar light is recycled in the infrared, then
$S_{\ast}\equiv S_{\rm bol} $.
Van der Veen (1989) gives,
for sources on the evolutionary sequence,
$S_{\rm OH} = S^2_{\ast}\,X^6_{21}$, where $X_{21}\equiv\,S_{25}/S_{12}$.
We find that this relation is not exactly right and also that
the scatter around it is very large.
Rather, for sources close to the evolutionary sequence (\Fg\VDVA ), 

\eqnam\ESO
$$S_{\rm OH} = S^{1.5}_{\rm bol}\,X^{3.5}_{21} \eqno (3a)$$

and for the whole sample in general (\Fg\VDVB ) 

$$S_{\rm OH} = S^{1.5}_{\rm bol}\,X^{2}_{21} .
\eqno(\new b) $$

\noindent
However, the ``maser efficiency'', defined as $S_{\rm OH}/S_{\rm bol}$,
is proportional to $X^{6}_{21}$ for the very bluest sources (\Fg\BOL ), 
since of course 
$S_{\rm bol}$ itself varies strongly with $X_{21}$ (\Eqt\EBC ).
For the reddest sources, the maser efficiency is virtually constant.
With a scatter of about a decade 
in these log--log plots these relations cannot be used on a one--by--one
basis. We can conclude, however, that statistically
the OH flux density of circumstellar masers is a well--behaved function
of ``pumping'' ($S_{\rm bol}$) and ``optical depth'' ($R_{21}$).

Moreover, there is a very clear, almost linear correlation 
between the 60--\mum\ flux density and the OH flux density (\Fg\SOSI).
The 60--\mum\ band contains the 53--\mum\ pumping line,
which may explain the correlation with ``available pumping
photons'', even though the line would be too narrow to
influence the wide--band flux.
The red post--AGB sources have surprisingly
similar flux--density ratio to the evolutionary--sequence sources.
In \Fg\SOSI , we plot the slope of the relation found 
by Chen \etal(2001) for the ratio $S_{\rm OH}/S_{35}$, with the
IRAS flux density at the main pumping line at 35 \mum\ derived
by interpolation.
Not only is their relation not linear, it
is also less tight,
so it seems the $S_{60}$ is a more uniform indicator of pumping than
the interpolated $S_{35}$, possibly due to contaminating lines 
in the $S_{25}$ band.

\section{Conclusions}

An extensive comparison is given between several infrared and OH--maser
properties of an OH--selected sample of objects in the galactic plane.
All the OH specifics in
this paper apply only to the 1612--MHz maser line and may
be (quite) different when studying the main lines.
Four different selection criteria are presented for finding red
proto--planetary
nebulae with OH masers, that all give consistent selections. 
Most efficient in separating post--AGB stars
is the MSX two--colour diagram
for $[8-12]$ and $[15-21]$ that splits into four clear quadrants,
containing primarily 
PPNe, SFRs, AGB objects and post--AGB stars, respectively.
Two selection criteria are presented for finding early post--AGB OH stars 
and one for post--AGB stars of all ages.
A significant group of blue (likely) post--AGB stars with 60--\mum\ excess
remains unselected by all of the above criteria; these will be
discussed in Paper II.

None of the selected post--AGB groups 
show any true prevalence of irregular or
extreme--velocity outflows, rather about 70\% of their 
OH spectra are typical of regular, spherical,
thin--shell outflow with velocities between 9 \kms\ and 15 \kms .
However, the sources with very irregular spectra 
are located roughly in 
the bipolar--outflow IRAS region defined by Zijlstra \etal(2001).
The reddest PPN may have a slightly higher fraction of
genuine single--peaked sources.

Maser efficiency, defined as the ratio of OH luminosity to
bolometric luminosity, increases steeply 
with IRAS colour $R_{21}$ for blue sources, but is almost
constant for the reddest sources. 
Of all IRAS flux densities, the OH maser correlates most strongly,
and linearly, with the 60--\mum\ flux density.

\begin{acknowledgements}

MS thanks the Leidse Sterrewacht for kindly providing facilities
to finish the final version of this paper.

\end{acknowledgements}

\onecolumn 

\appendix
\newcount\fignumber \fignumber=1

\section{Spectra}

\figurenum{A1}
\begin{figure}
\fignam\SPI
\anfig
\psfig{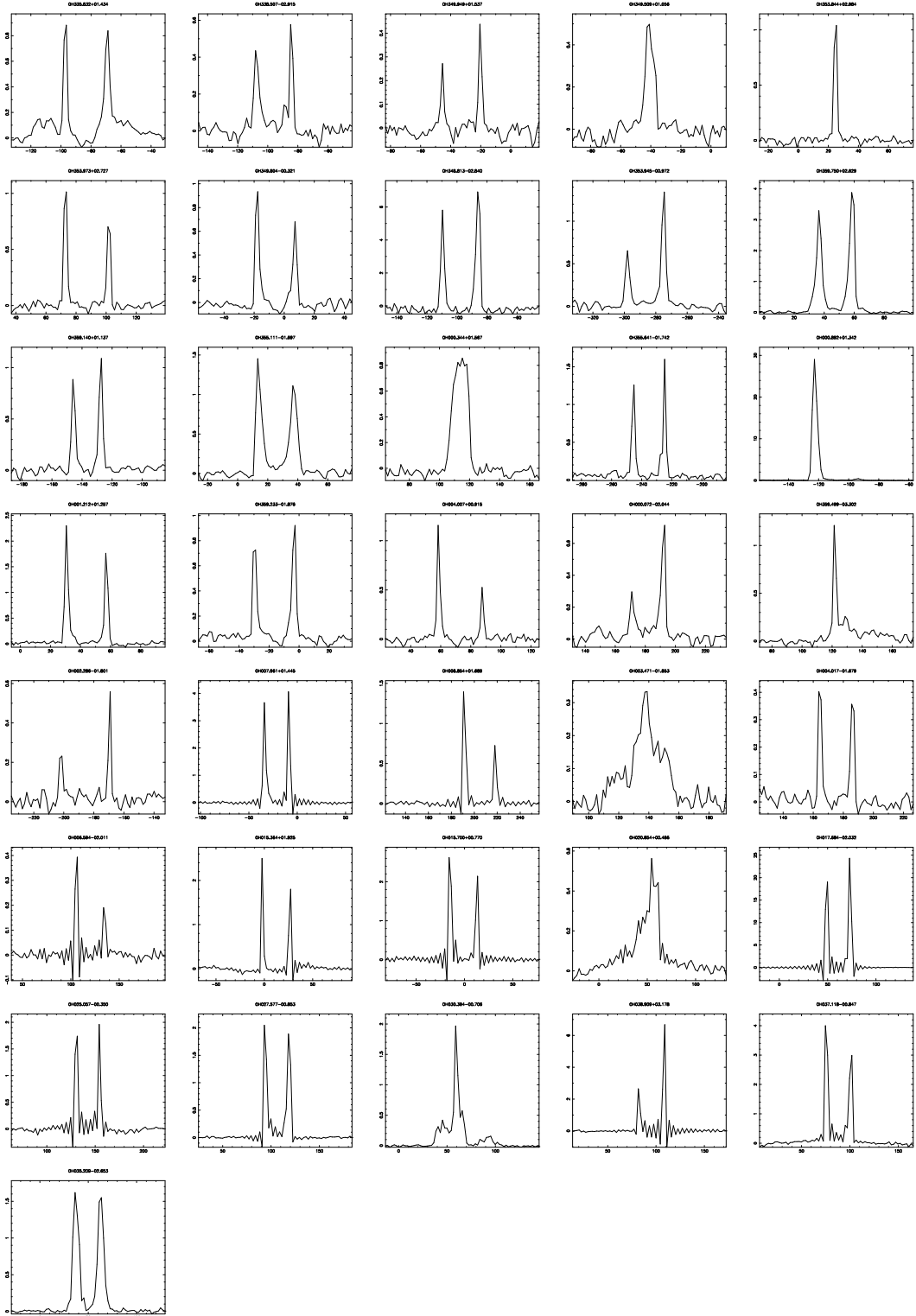}
\figcaption{
All sources with $R_{32}<$1.5 and $R_{21}>$1.4 (item 1 in \Sct 5.1).
Note that all spectra in this appendix have flux densities
that are NOT corrected for primary--beam dilution. In 
the text, quoted flux densities ARE corrected.
}
\end{figure}

\vfill
\eject

\figurenum{A2}
\begin{figure}
\fignam\SPM
\anfig
\psfig{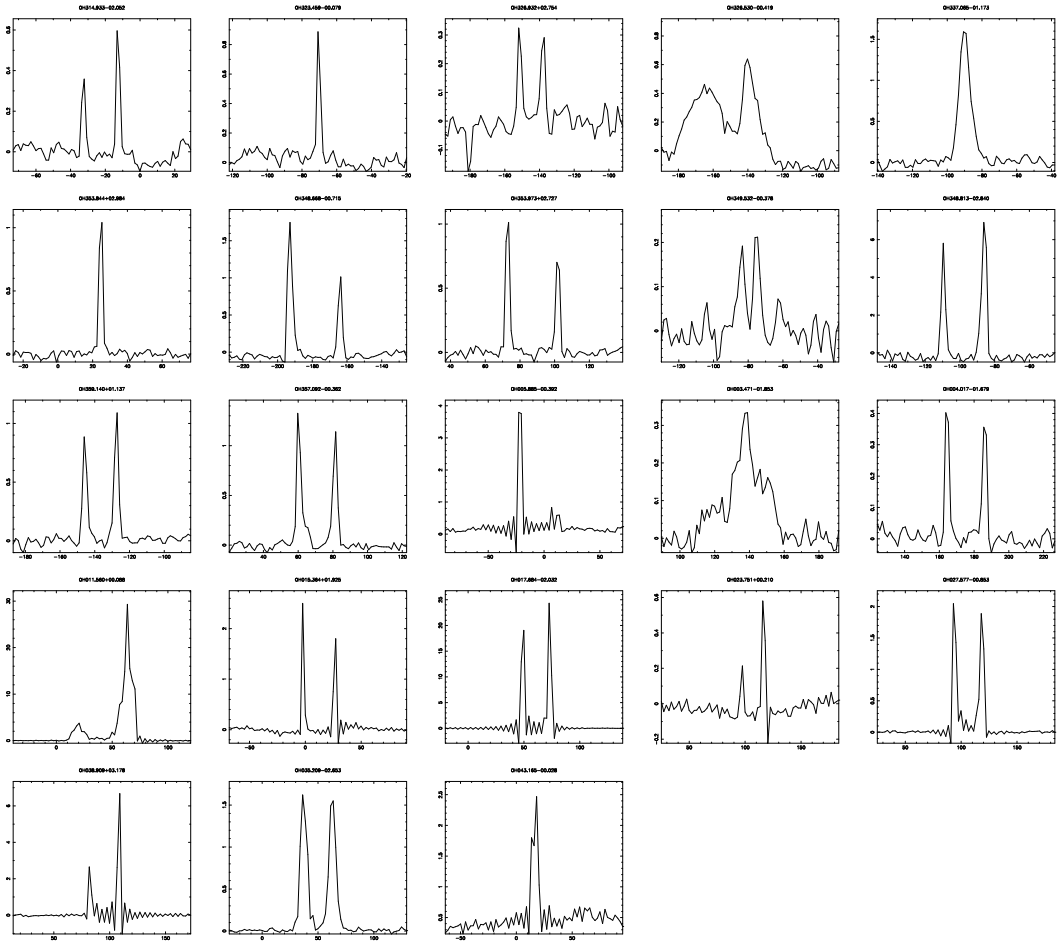}
\figcaption{
All sources with [8-12]$>$ 0.9 and [15-21]$>$0.7 (item 2 in \Sct 5.1).
}
\end{figure}

\vfill
\eject

\figurenum{A3}
\begin{figure}
\fignam\SPH
\anfig
\psfig{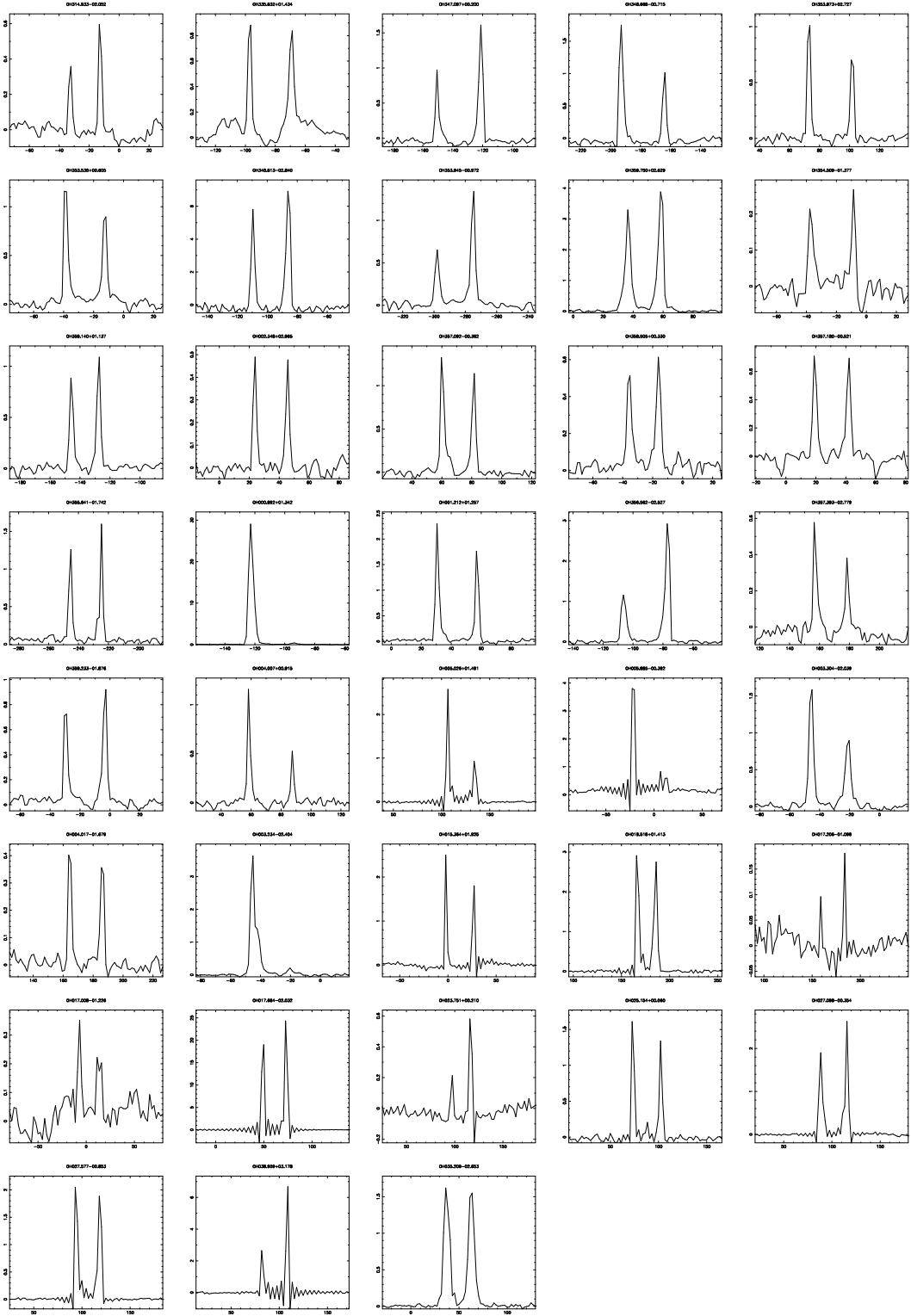}
\figcaption{
All sources with [15-21]$>$0.7 and 9\kms $V_{}<$15 \kms\ 
and sum of the blue-- and red--shifted OH peak flux densities
larger than 1 Jy (item 3 in \Sct 5.1, \Fg\HFA ).
This gives [12]-[25]$>$3 and [8-12]$>$0.9 .
}
\end{figure}

\vfill
\eject

\figurenum{A4}
\begin{figure}
\fignam\SBM
\anfig
\psfig{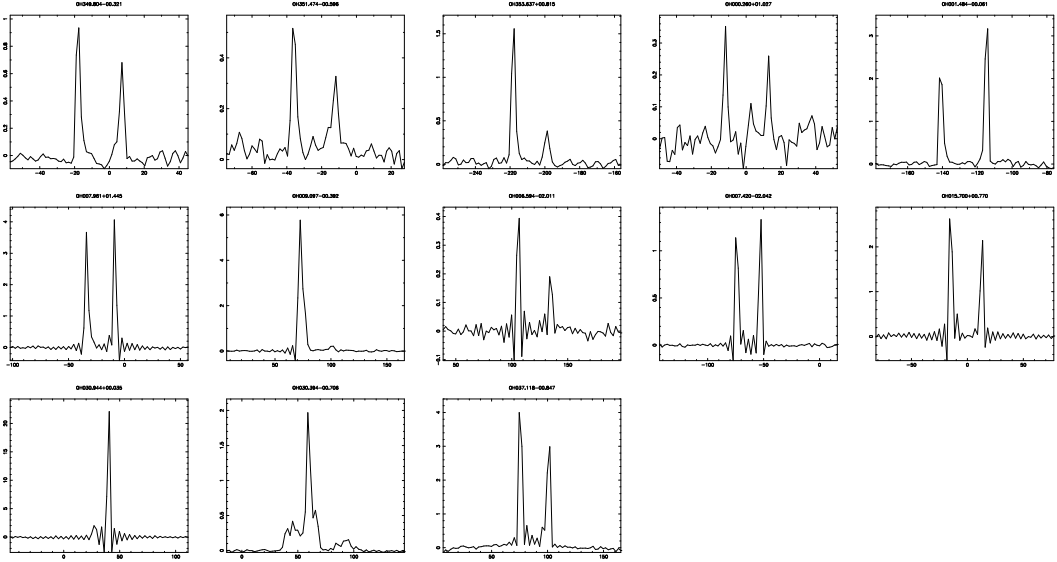}
\figcaption{
Sources with  [15-21]$<$0.7 and [8-12]$>$0.9 (item 5 in \Sct 5.1).
Second last source has [15-21]$\sim$0.7 though
}
\end{figure}

\vfill
\eject

\figurenum{A5}
\begin{figure}
\fignam\SCI
\anfig
\psfig{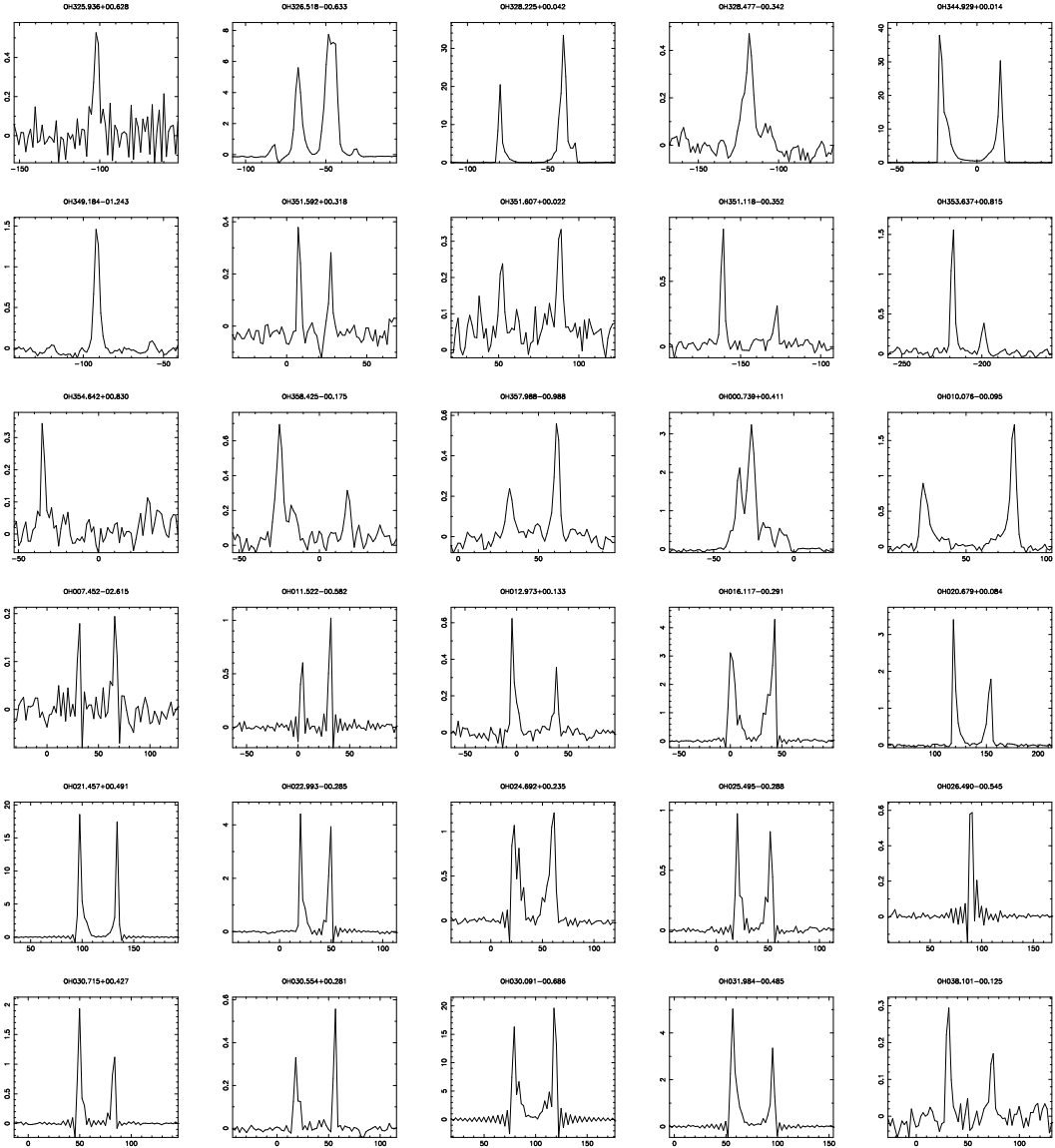}
\figcaption{
All sources with $R_{32}>-$0.2 and $R_{21}>$0.2 to the left of
the evolutionary sequence in the IRAS two--colour 
diagram (item 8 in \Sct 5.1, \Fg\IRSA ).
}
\end{figure}

\vfill
\eject

\figurenum{A6}
\begin{figure}
\fignam\SPS
\anfig
\psfig{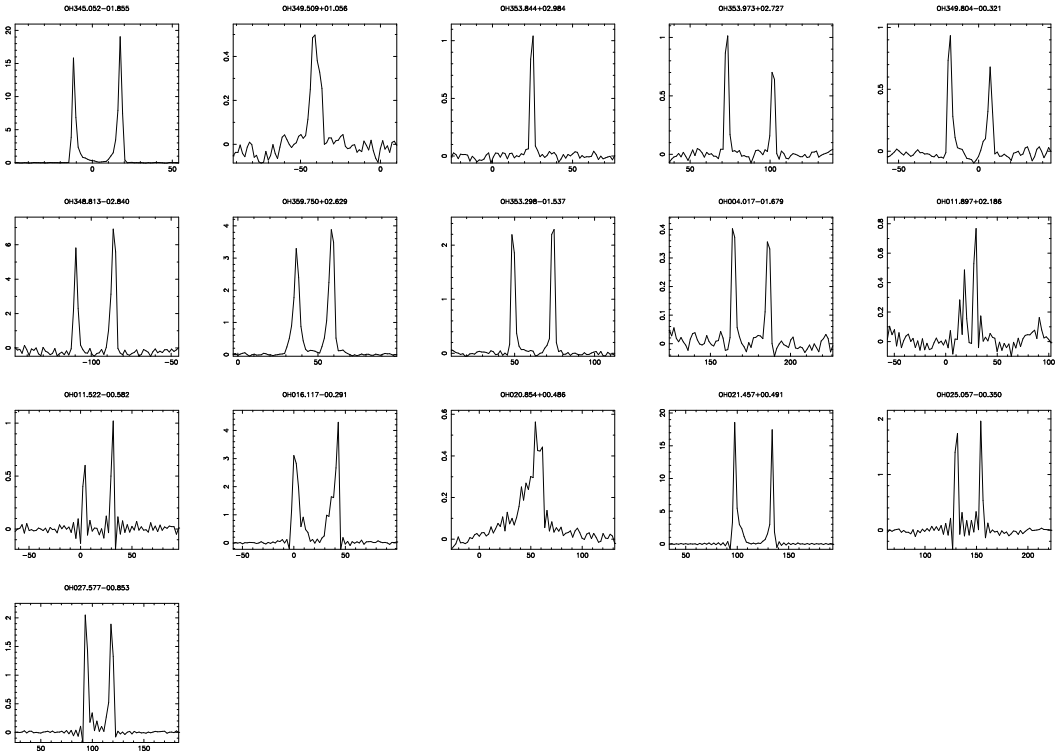}
\figcaption{
All sources with a SIMBAD association of PN within
5\arcsec\ (\Sct 5.2).
}
\end{figure}

\vfill
\eject

\figurenum{A7}
\begin{figure}
\fignam\SIS
\anfig
\psfig{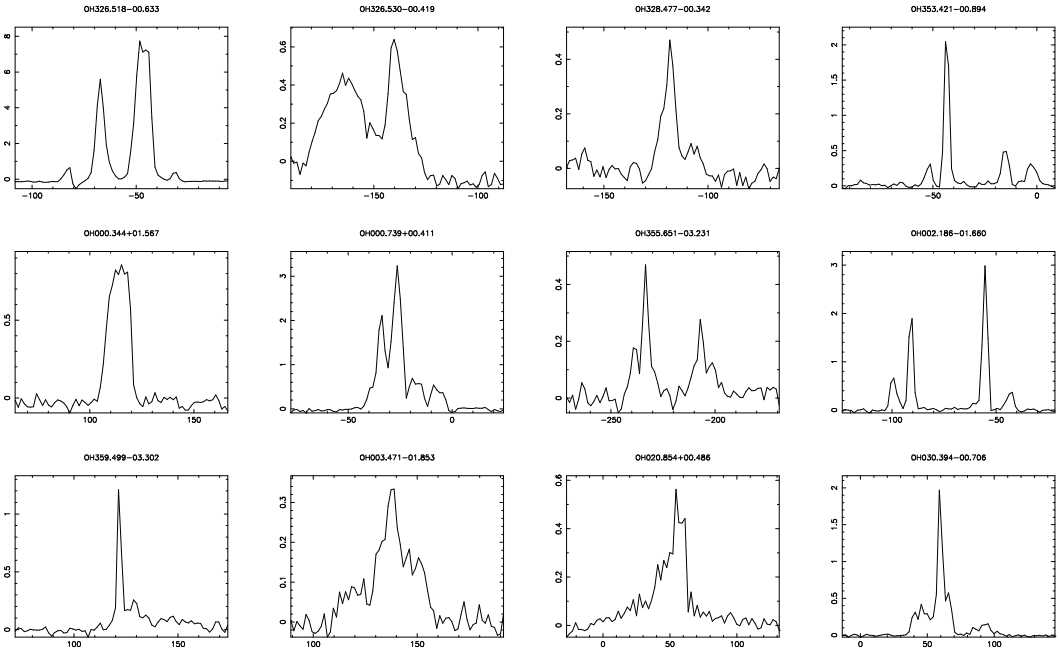}
\figcaption{
All sources with ``irregular spectra'', defined 
by $w_{20} >> \Delta V $ (\Sct 3). 
}
\end{figure}

\vfill
\eject

\figurenum{A8}
\begin{figure}
\fignam\SES
\anfig
\psfig{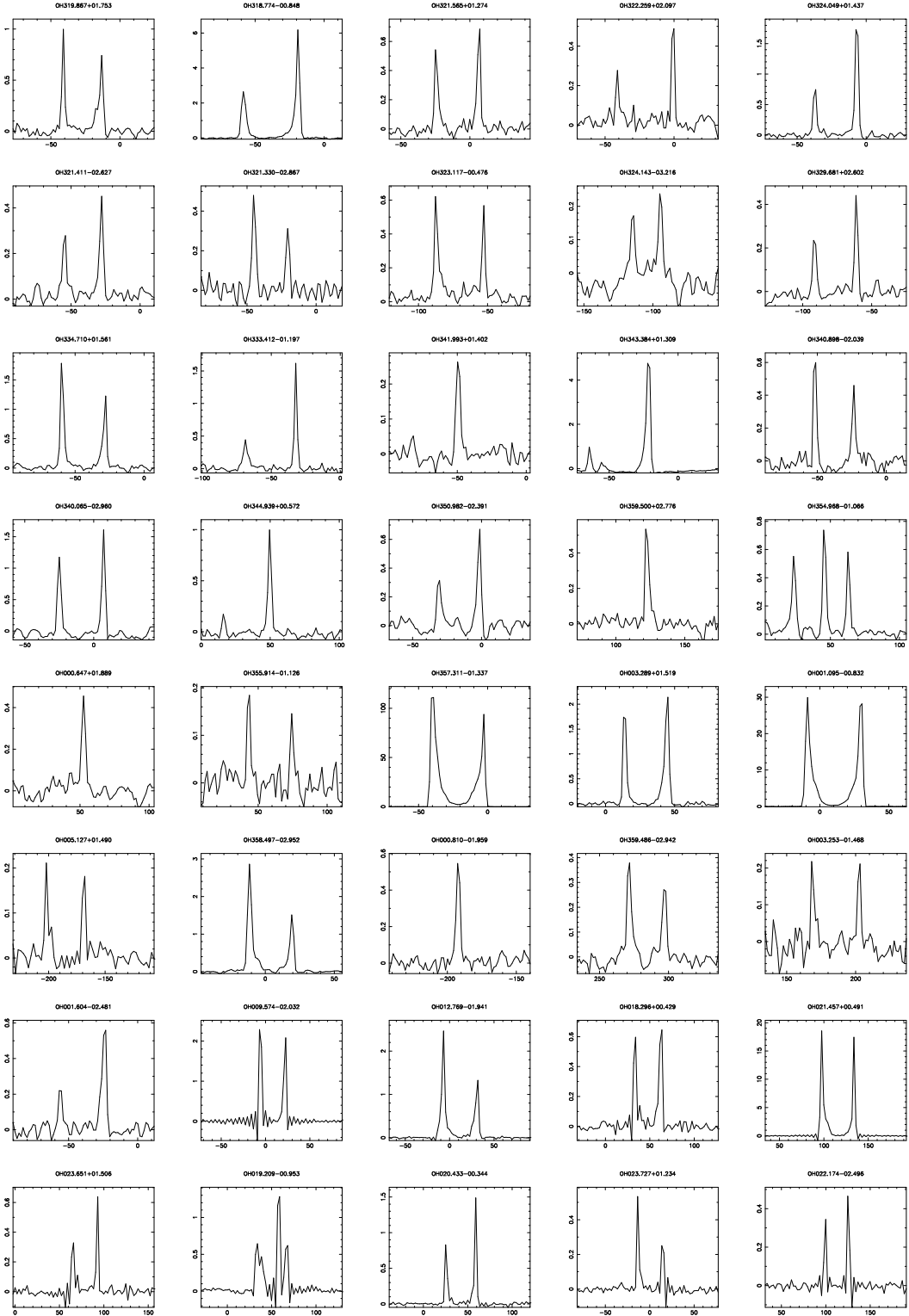}
\figcaption{
Thermally--pulsing AGB sources, on the IRAS evolutionary sequence
(solid curve in \Fg\IRSA ).
}
\end{figure}

\end{document}